# Love before Sex[1]

Wan Ahmad Tajuddin Wan Abdullah[2]
*Jabatan Fizik, Universiti Malaya*
*50603 Kuala Lumpur*

Much has been debated about the benefit of sexual over asexual reproduction in terms of evolutionary fitness. Here we focus on the advantage that may be brought about by the process of mating, where the choosing of mates contributes to the increase in fitness in a constructive way. We carry out computer simulations of such mating systems and investigate, on one hand, how mate phenotypes contribute to offspring fitness, and, on the other hand, how selection affects mate phenotypes. We discuss how helpful such a mechanism may be in determining trajectories on rugged energy landscapes leading to global optimum.

## Introduction

The study of complex systems has emerged as an important frontier in physics. One defining characteristic of complex systems is the ability to adapt to a changing environment. This adaptibility has been associated with the process of optimization, and more basically, as the search for the maximum or minimum of some resp. fitness or cost function. This of course parallels the thermodynamic search for stable states in materials, including in complex structures like amorphous and glassy ones, where the lowest valleys of the energy landscape in the multivariate configuration space. In some complex systems like biological and ecological ones, the optimization lanscape can be very complex, with multiple valleys, and can be changing in time.

Evolutionary or genetic algorithms [1] (see Figure 1 for a schematic description) have been proposed to model biological evolutionary adaptation, and provide a good parallel methodology for optimum searching, but would be rather inflexible to landscape changes at time rates faster than the typical relaxation time of the evolutionary process. Table 1 shows the other processes that exist in addition to evolution that provides for optimization at these shorter timescales. Learning and memory has been studied extensively in the literature of both backpropagation feed-forward [2] and symmetric Little-Hopfield [3] neural networks, while culture has been modeled [4] in terms of direct transference of (learnt) knowledge between agents.

---

2 http://fizik.um.edu.my/cgi-bin/hitkat?wat

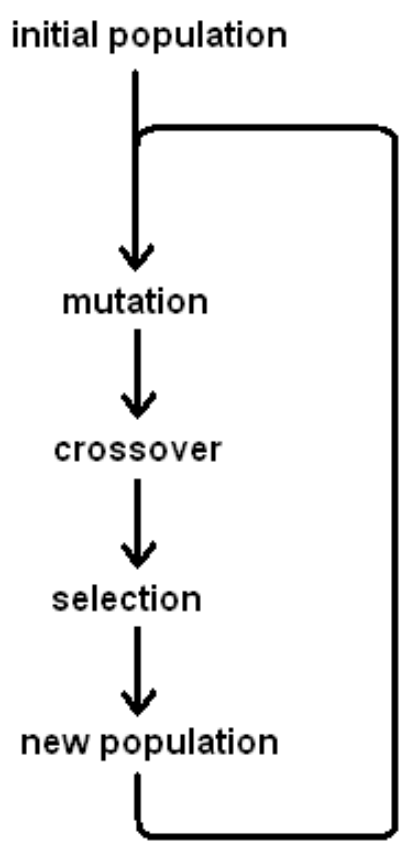

**Figure 1** *Evolutionary algorithm*

| Optimization process | Timescale |
|---|---|
| evolution | many generations |
| culture | several generations |
| learning | single generation |
| memory | fraction of single generation |

**Table 1** *Optimization processes at different timescales*

Given the interplay between evolution and these other processes it is interesting to ask whether there may be other processes, maybe implicit, that improves optimization. It is thus the object of this paper to examine if mate selection ("love") in evolution with sexual reproduction provides any advantage.

This study can also be viewed as a investigation of an aspect of *meta-evolution*, i.e. the exploration of mechanisms/operations/components of evolution which enhances it. This embraces for example the asexual/bisexual reproduction debate [5], or in general the *n*-sexual question [6], mating strategies (in bisexual systems), etc.

Since we are subjecting mate selection in this paper to the underlying evolutionary process as well, this study can also be seen as an investigation into *self-evolution*, i.e. evolution with its parameters subjected to itself. The evolution of evolvability itself can be studied [7]. Previously [8], we studied the evolution to the optimum 'number of sexes' (i.e. number of parties to crossover). We have also studied [9] systems where mating frequencies and lifespans are themselves subjected to evolutionary optimization; this is interesting because intuitively, on one hand increase in fitness would mean longer lifespans but on the other hand shorter lifespans would mean faster optimization. Along the same lines, heterogeneity and mating preferences can be described by parameters to the evolutionary process to which they are themselves subject.

We investigate the validity of the intuitive notion that mate selection would accelerate evolutionary optimization, and the effect of evolutionary selection on mate selection, using a model in which (parts of) genes of individuals determine mate selection. We describe this model below, and subsequently report on the computer simulations carried out and the results.

## The Model

We now describe our model. We have a population of a fixed size with a common mating rate. Each member of the population has a fixed common lifetime, given as a certain number of mating cycles, and is described by a binary string for genes. A part of the genes, the 'physical' genes, determine the fitness of the individual, while the rest, the 'love' genes, do not contribute to the fitness but are used to choose mates: the suitor with the physical genes most similar to her love genes get preferred. Mate selection is done through choosing the best from a certain fixed number of random suitors.

We choose a polynomial form for the fitness function. Intuitively, this seems to be the most plausible form for it. We take it to be a sum of first, second, ... order randomly-weighted conjunctions of gene bits,

$$F = \alpha_1 b_{i1} + \alpha_2 b_{i2} + \ldots + \beta_1 b_{j11}\, b_{j12} + \beta_2 b_{j21}\, b_{j22} + \ldots + \gamma_1 b_{k11}\, b_{k12}\, b_{k13} + \ldots,$$

where $b_{i1}$, $b_{j11}$, etc are gene bits and $\alpha_1$, $\beta_1$, $\gamma_1$, etc are random numbers between 0 and 1. Without loss of generality (modulo permutation of bit positions), we order the bits according to their order of occurrence in the sum. Due to the implied bit position permutation, we choose to implement crossover by picking randomly from either parent gene bits bit position by bit position (and not by exchanging random lengths of gene string portions as is the usual practice).

## The Simulation and Results

We carried out simulations of the model for a population size of 30 individuals initially and maximum at any one time, each with 18 bits of physical genes: 3 bits contributing to 3 first-order terms, 6 to 3 second- order terms, and 9 to 3 third-order terms in the fitness function. The maximum fitness is then $(\alpha_1 + \alpha_2 + \alpha_3 + \beta_1 + \beta_2 + \beta_3 + \gamma_1 + \gamma_2 + \gamma_3)$, while a totally random state, taking means of weights and bits, would have a fitness ratio of $(3\times0.5 + 3\times0.25 + 3\times0.125)/9$, or 0.29, to that of the maximum.

The individual lifespans are fixed at 4 mating cycles each while all 30 (or whatever the current number is) individuals mate in a mating cycle. The fittest individuals, after mutation and mating crossover, form the new offsprings as well as parents still within their respective lifespans, survive to the next round. Mate selection is carried out by selecting the best from among 5 random suitors, best being the one with its physicals genes most similar to the love genes of the first individual. Mutation changes random bits in individual genes (physical and love) by a specified rate.

The simulations involve 100 random trials for each set of parameters and conditions tested, giving statistical errors on the means to be smaller than the data points.

Since know what the respective optimal fitnesses should be from definition, we can compare convergence rates for the various parameters. For a random selection of fitness gene bits, the fitness is on average around 0.3 of the optimum fitness with our choice of the fitness function.

Figure 2 shows the result of a comparison between different mutation rates given the same mating rate. Mutations are analogous to noise, and higher mutations show slower convergence. Whereas noise can aid convergence, it seems that crossover is adequate. However, although zero mutation shows faster convergence initially, it somehow could not really get to the correct optimum state in the long run, and some slight mutation seems necessary. We take the mutation rate of 0.01 as better and use this value in further simulations.

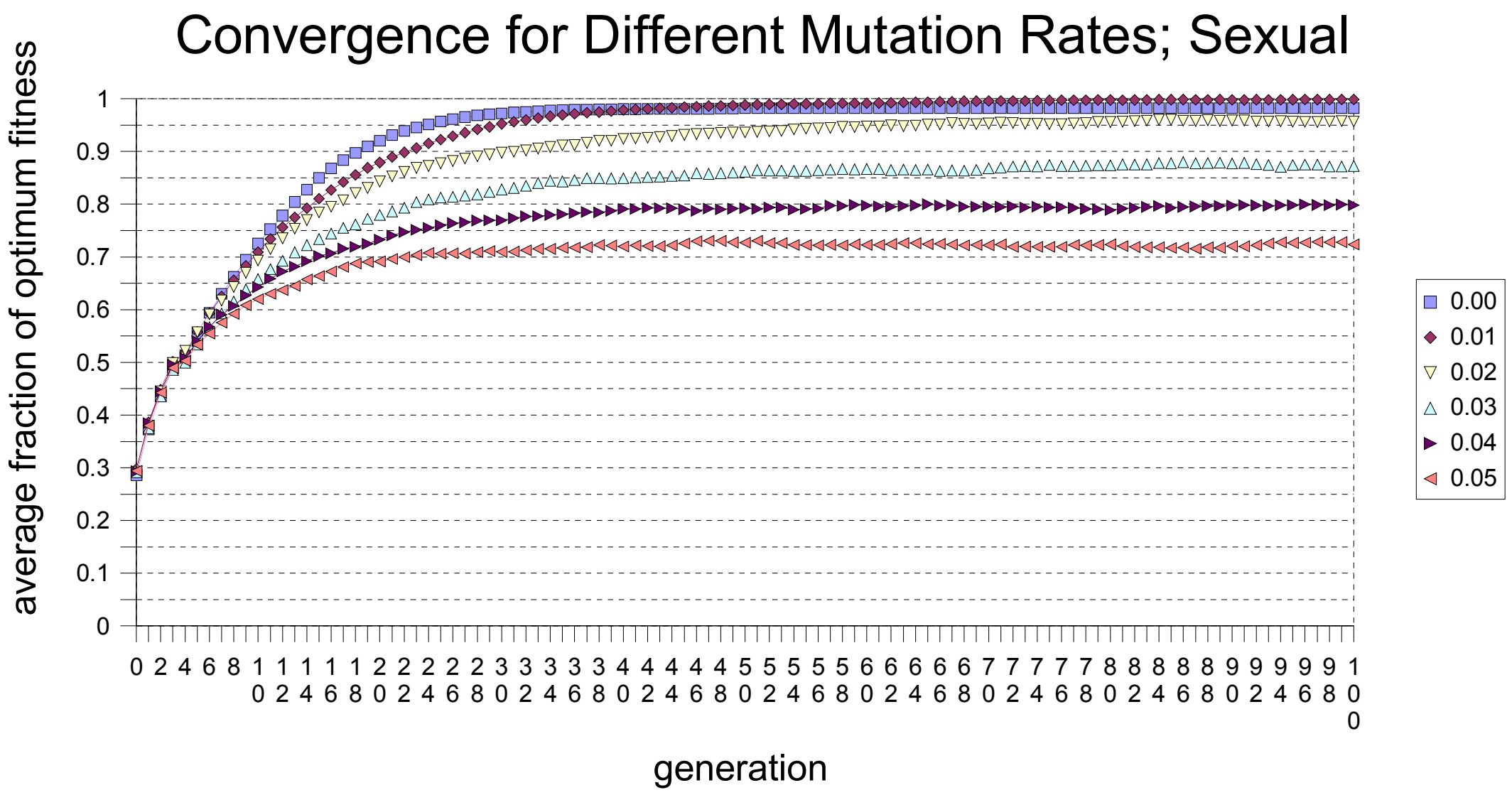

**Figure 2**

In Figure 3, we allow the mating rates to vary and compare convergence rates for the different mating rates. The mating rate is given by the percentage of individuals from the maximum 30 which mate in each mating cycle. As may be expected, convergence is faster for a higher mating rate. This gives support to the sexual case in the sexual-asexual debate. The 40% mating rate curve is interesting as it initially converges but then falls back onto the ‘random’ line of 0.29, indicating complete information absence.

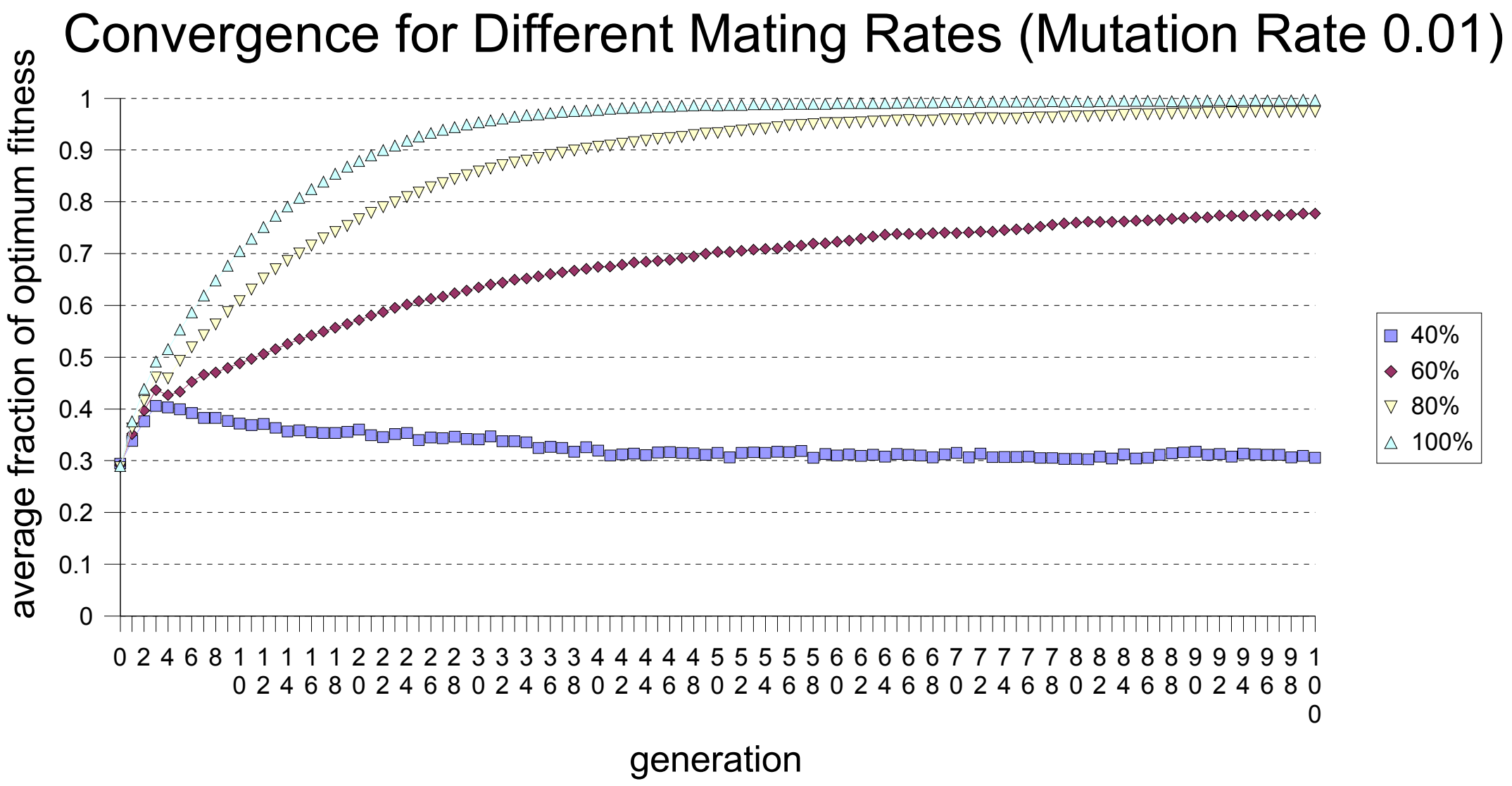

**Figure 3**

We now investigate the effect, if any, of 'love' or mate selection. Figure 4 shows the convergence behaviours of systems with and without mate selection, and for the latter, with selection between different numbers of suitors. Surprisingly, there is no indication of any advantage in mate selection! One may argue that with 100% mating, mate selection could have minimal effect since every individual mates anyway. However, Figure 5 also shows similarly negative results for 60% mating.

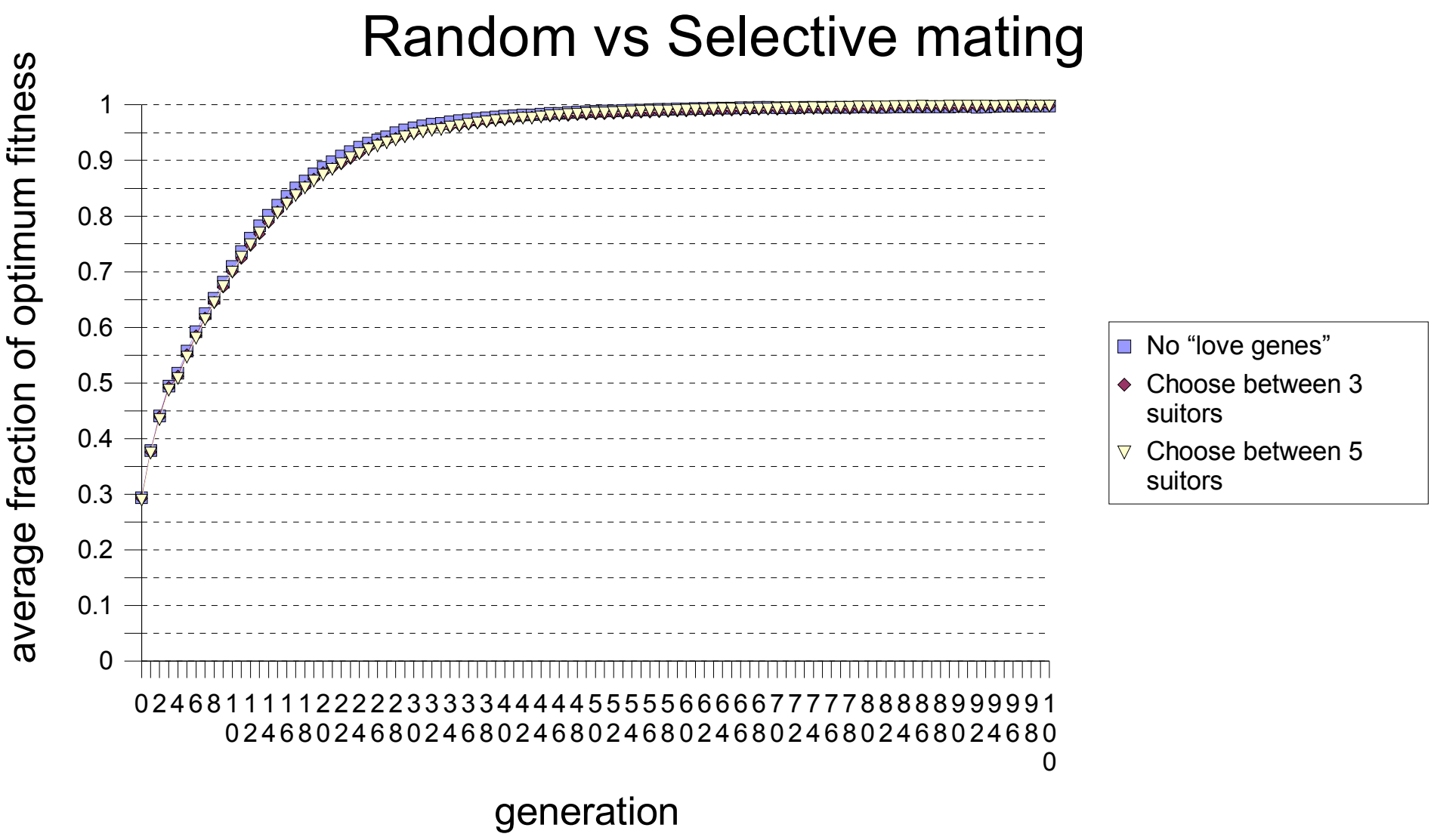

**Figure 4**

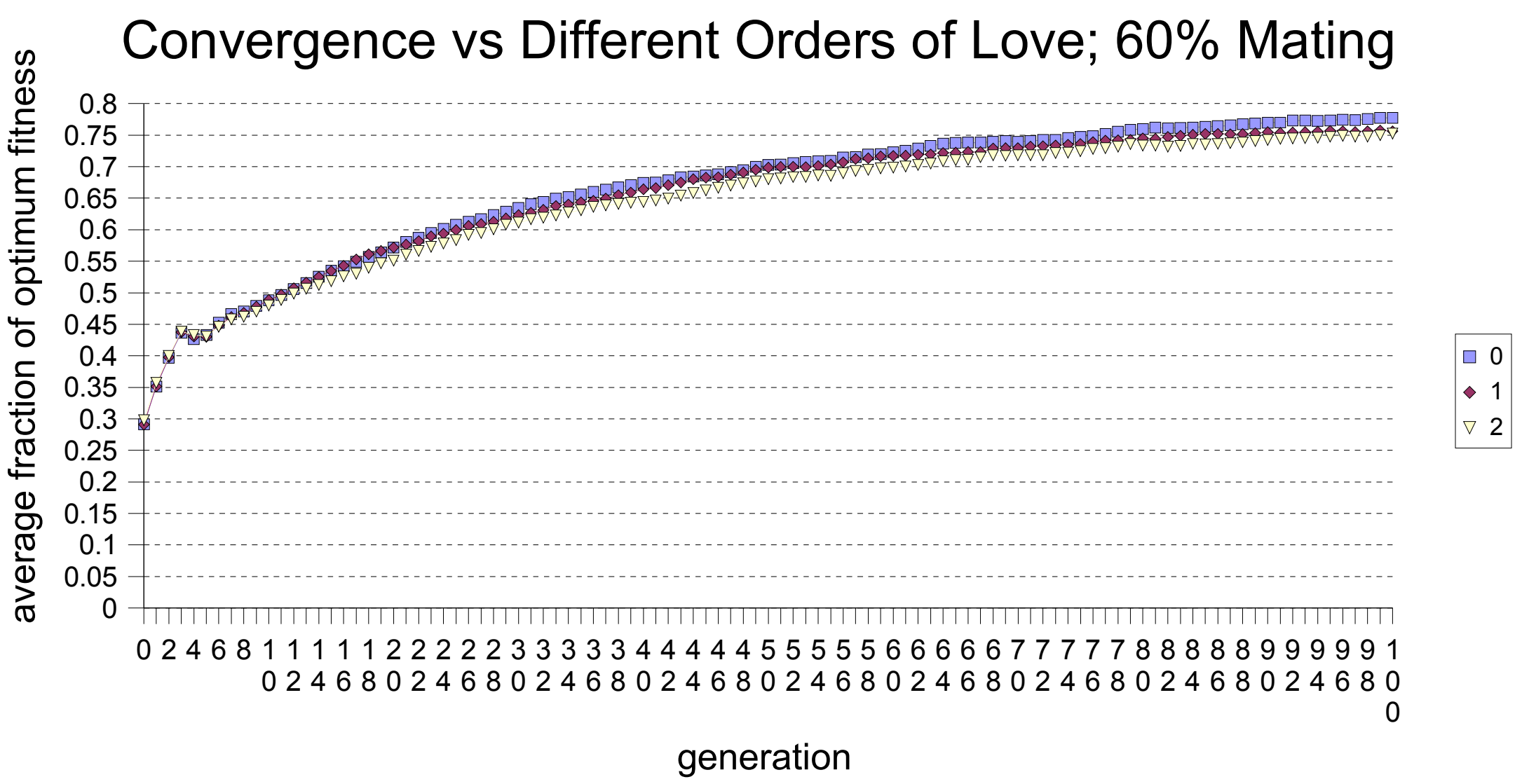

**Figure 5**

We can also look at the effect of ‘higher-order love’ – we can include second-order love genes which act with respect to the physical+(first-order )love genes in a similar way the love genes act with respect to physical genes, and then third-order love genes similarly, and so on. However since first-order love itself has no effect on convergence rates, we do not expect any from the other orders, and this is proven by the simulation results in Figure 6.

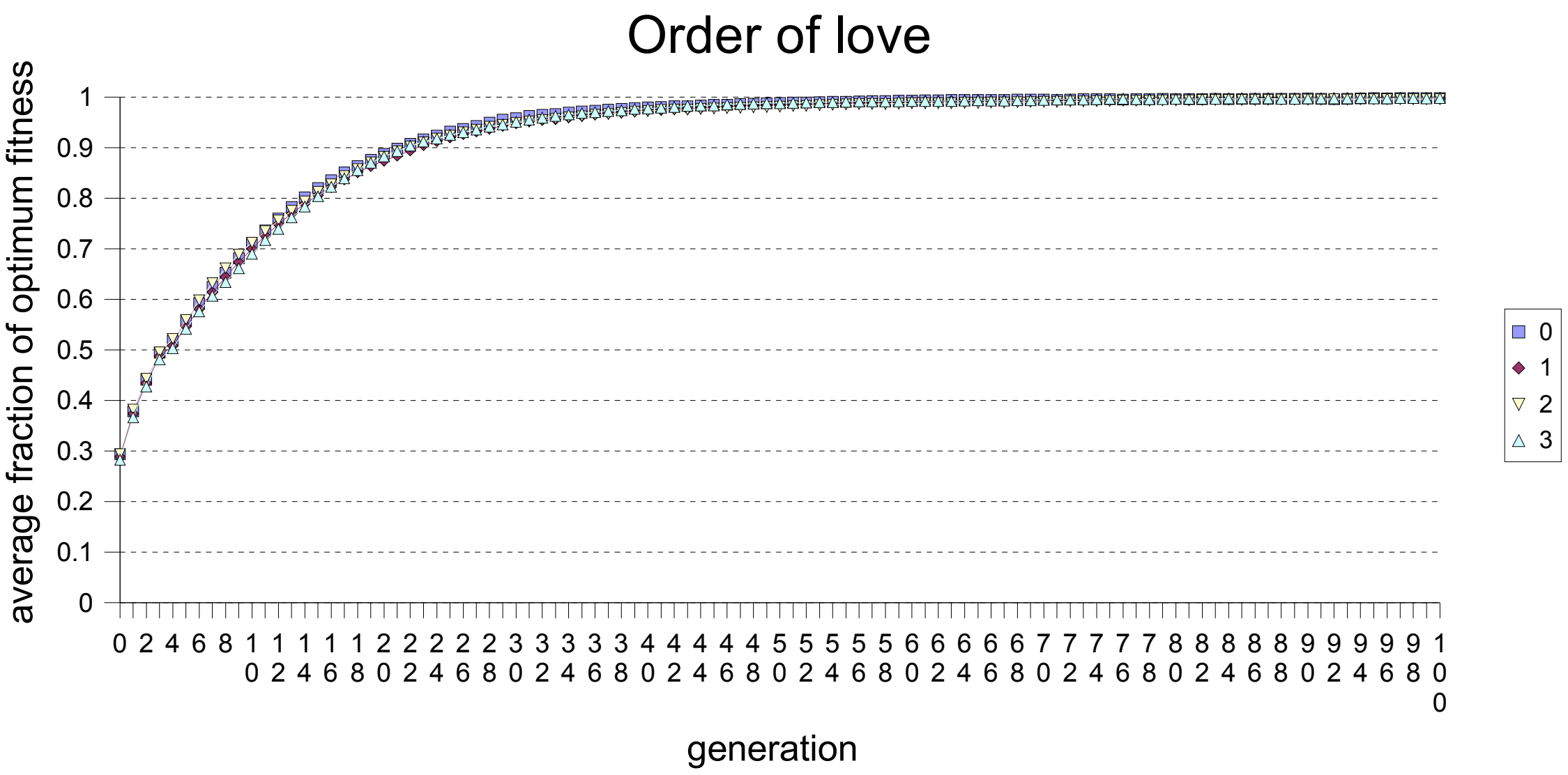

**Figure 6**

If we now look at how the love genes change with the evolutionary selection, we obtain results as given in Figure 7. It has to be remembered that only the physical genes are subjected directly to selection, and love genes, while they change with crossover and mutation, are not considered in selection. It is thus rather counter-intuitive to find that the evolutionary process enhances mate selectiveness, as shown by the decreasing dissimilarity between the love genes and the physical genes that they respectively select.

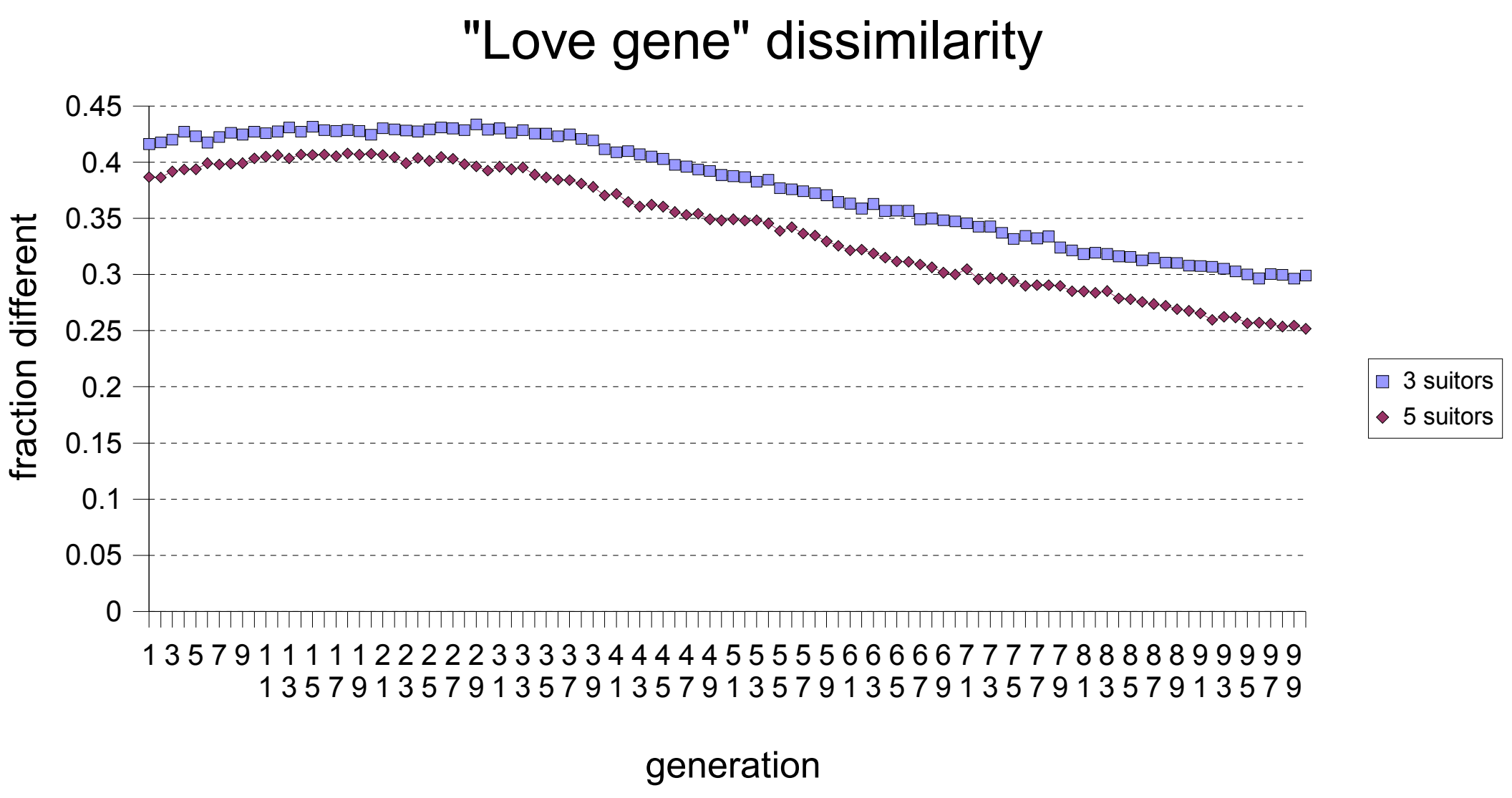

**Figure 7**

## Conclusions

We have proposed an evolutionary model with which we studied the effect of mate selection on evolutionary convergence in sexual populations. The model indicates the advantage of sexual reproduction from the increased convergence with increased mating rate. However, we find no evidence for increased convergence with mate selection, and this is also as such for the several orders of love investigated.

On the other hand, the evolutionary process enhances mate selection in the sense that more suited suitors are selected. Love, then, does not enhance sex, but rather, sex enhances love!